\documentclass[a4paper,12pt]{article} 
\usepackage{graphicx}

\newcommand{\beq}{\begin{eqnarray}}
\newcommand{\eeq}{\end{eqnarray}}

\begin{document}
\pagestyle{plain}

\title{
{\Large \bf 
Status of the ATF extraction line laser-wire.
}
}
 
\author{
{\small \bf Nicolas Delerue, Fred Gannaway, David Howell } \\
{\small \em John Adams Institute at the University of Oxford,} \\ {\em \small Keble Road, OX1 3RH, Oxford, United Kingdom } \vspace{.5cm} \\ 
{\small \bf Grahame Blair, Gary Boorman, Chafik Driouichi  } \\
{\small \em John Adams Institute at Royal Holloway, University of London,} \\ {\em \small Egham, Surrey TW20 0EX, United Kingdom }  \vspace{.5cm} \\ 
{\small \bf Stewart Boogert } \\
{\small \em Department of Physics \& Astronomy, University College London,} \\ {\em \small Gower Street, London, WC1E 6BT United Kingdom }  \vspace{.5cm} \\ 
{\small \bf Alexander Aryshev, Pavel Karataev, Nobuhiro Terunuma, Junji Urakawa } \\
{\small \em KEK, 1-1 Oho, 305-0801 Tsukuba Science City, Japan }  \vspace{.5cm} \\ 
{\small \bf Axel Brachmann, Joe Frisch, Marc Ross } \\ 
{\small \em SLAC, 2575 Sand Hill Road, Menlo Park, CA 94025, USA } \\
}

\date{
December 2005
}

\maketitle

%
\begin{center}
{\bf 
A new laser-wire is being installed in the extraction line of the ATF at KEK. This device aims at demonstrating that laser-wires can be used to measure micrometre scale beam size.
} \\ 
\end{center}

\section{Overview of the ATF extraction line laser-wire}

The Accelerator Test Facility (ATF) at KEK offers a very low emittance  micrometre sized beam of 1.28 GeV electrons close from the conditions expected at the International Linear Collider (ILC). The ATF should soon be able to provide a bunch spacing similar to that of the ILC. This accelerator being a test facility it is possible to have dedicated machine time with custom beam optics for our tests. This make the ATF an ideal place to test a new beam diagnostic device such as a laser-wire to measure micrometre sized beams.

For this laser-wire installation KEK has provided a high power green laser that can deliver up to 1 GW at 532nm (this laser was previously used by the polarised positron experiment).

\section{Laser system}

A mode-locked seed laser (Nd:VAN) produced by Time-Bandwidth (GE-100-XHP series) is used to deliver a 600mW train of light pulses with a wavelength of 1064 nm and a repetition rate of 357 MHz. 
The Nd:VAN crystal is diode pumped and a semiconductor saturable absorber (SESAM)\cite{sesam1,sesam2} is used to start and stabilise the pulse forming process.
The SESAM mirror forms one end of the laser cavity. It is mounted on a translation stage so that the length of the cavity can be adjusted by a picomotor and a piezo-electric crystal to cancel the effects caused by changes in ambient conditions. 
A photo-diode is used to monitor the output of the cavity. A timing stabiliser (Time Bandwidth CLX-1100)\cite{timing1,timing2} is used to phase-lock the signal from the photo-diode  to the ATF 357 MHz RF signal (This is achieved by adjusting the cavity length).

The pulse train produced by the seed laser is injected in a regenerative amplifier custom-built by Positive Light (model RGN). This system amplifies one of the nanoJoule seed pulses to approximately 600 milliJoules. 
A Faraday isolator is used to prevent reflected pulses damaging the seed laser.
Two Pockels cells are used to select the pulse that will be amplified. The first Pockels cell is located before the Faraday isolator, it is used to chop the end of the pulse. The second Pockels cell is located inside the amplifying cavity. When this second Pockels cell is deactivated the pulses do not reach the amplifying rod in the cavity. When this Pockels cell is switched on, it will trap one pulse inside the cavity. This pulse is reflected through an Nd:YAG amplifying rod and bounces back and forth in the cavity. After a number of pulse round trips in the cavity (usually 10-15) a third Pockels cell is switched on, this changes the polarisation of the pulse inside the cavity and this pulse is extracted by a polariser making a 45 degrees angle with the beam trajectory.

After extraction the pulse passes a second Faraday isolator that protects the cavity. It is then transmitted through a spatial filter and then through 2 Nd:YAG linear amplifier that amplify the pulse by a factor of 10 each, bringing it to more than 600 mJ. Before the exit of the laser system a KD*P crystal doubles the frequency of the laser pulse, bringing its wavelength to 532 nm with an efficiency close to 50\% at full power. At this stage the pulses have a length of 200-300~ps.

The timing of the Pockels cells is critical to ensure that one and only one pulse is amplified and to improve the contrast ratio of the pulse by removing any pre and post pulses. This timing is controlled by a signal and delay generator (SDG-II from Positive Light). A photo-diode monitors the signal in the amplifying cavity. A second photo-diode, located after the spatial filter, monitors the signal extracted from the cavity. These two photo-diodes are used to tune the timing of the 3 Pockels cells and thus improve the extracted signal.

The amplifying rod located inside the cavity and the two linear amplifiers are optically pumped by flash lamps. The signal used to trigger the flash is taken from the ATF extraction kicker charge signal. The kicker fire signal is used to trigger the SDG and thus the amplified laser pulse.

\section{Laser delivery optics and focusing lens}

The laser system described above is located on top of the ATF shielding. A set of mirrors are used to bring the laser light from the output of the amplifier to the focusing light.
 The total length travelled by the light is of the order of 10 meters. A scanning system and diagnostic tools such as photo-diodes and camera will be installed along this path.

A F\#/2 lens will be used to focus the laser light onto a very small spot\cite{bdir_nd}. This lens is made of 3 elements: the first element has an aspheric surface and a spheric one. The second element has two spheric surfaces. The last element is flat and is used as a window to allow the laser light to enter the beam pipe. All these element are made of top-quality fused silica.

\begin{figure}[htbp]
\begin{center}
\includegraphics[height=8cm]{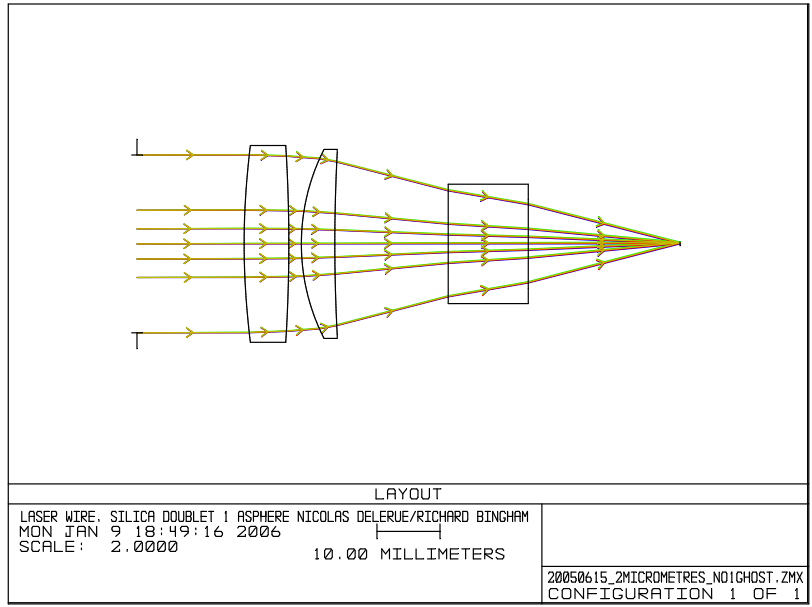}
\caption{Layout of the F\#/2 lens designed for the ATF extraction line laser-wire \label{fig:lens}  }
\end{center}
\end{figure}

Beam dynamics and mechanical considerations require the inner side of the window to be more than 20~mm away from the IP which must be roughly in the centre of the beam pipe, in our design this inner surface of the window is 24~mm away  from the IP. The window has a thickness of 12.7~mm. The position of the two other elements is constrained by mechanical and cost consideration: to allow the sealing of the window these two elements must be more than 14~mm away from the window but they must be kept as close as possible to the window to limit their size (and hence their cost). In our design one of these elements is located 18~mm away from the window and has a thickness of 5.3~mm. The second element (aspheric) is located 2~mm further away and has a thickness of 7~mm. The layout of this lens is shown in figure~\ref{fig:lens}.

\begin{figure}[htbp]
\begin{center}
\includegraphics[height=6cm]{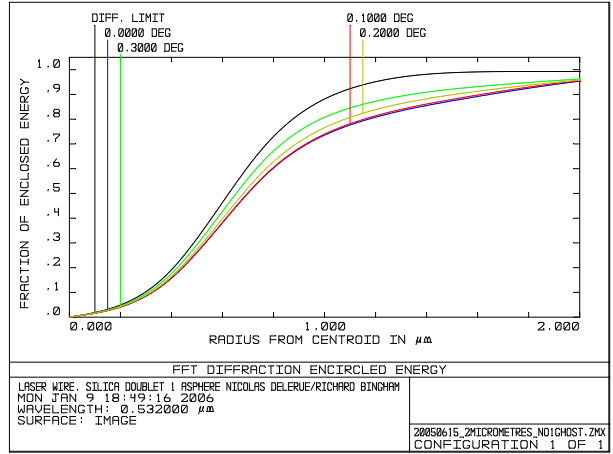}
\caption{Fraction of enclosed energy in a given circle as a function of the radius of this circle from the centre of the beam spot. The upper (black) line shows the diffraction limit, the 4~other lines show the real value when the incoming laser beam has a tilt of 0 degree (blue), 0.1 degree (red), 0.2 degrees (yellow) and 0.3 degrees (green). \label{fig:lens_diff}  }
\end{center}
\end{figure}

The shape of these elements has been optimised using the Optical Design Software ZEMAX\cite{zemax} to focus the beam to a spot size as close as possible from the diffraction limit as shown in figure~\ref{fig:lens_diff}. As the laser beam will be scanning the electron beam, the optimisation has been done for 4 different tilt angles: 0 degree (no tilt), 0.1 degree, 0.2 degree and 0.3 degree and one of the goals of the optimisation was to keep the size of the laser spot in each of these configurations a identical as possible as can be seen in figure~\ref{fig:lens_diff}.

\begin{figure}[htbp]
\begin{center}
\begin{tabular}{ccc}
\includegraphics[height=6cm]{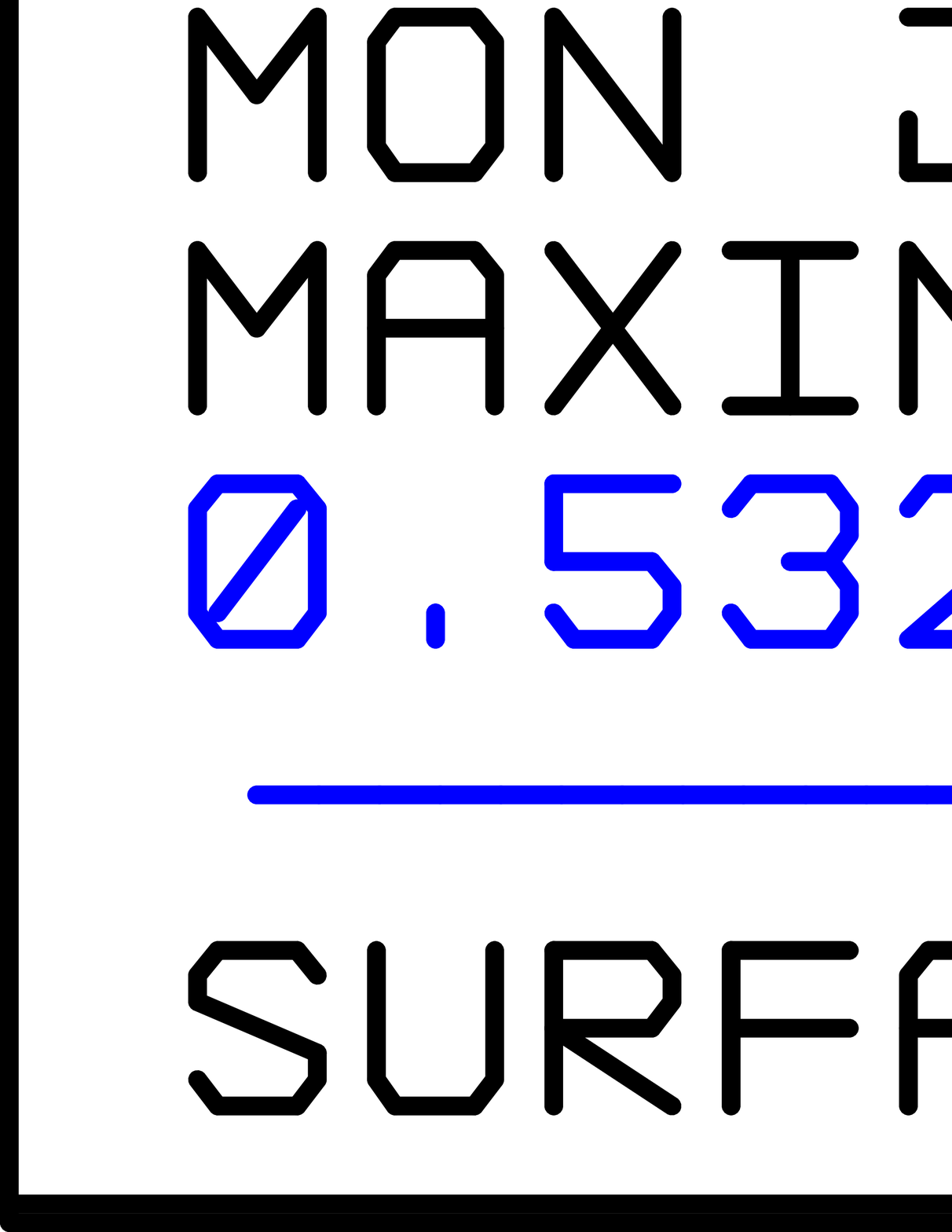} & \hspace{.15cm} & \includegraphics[height=6cm]{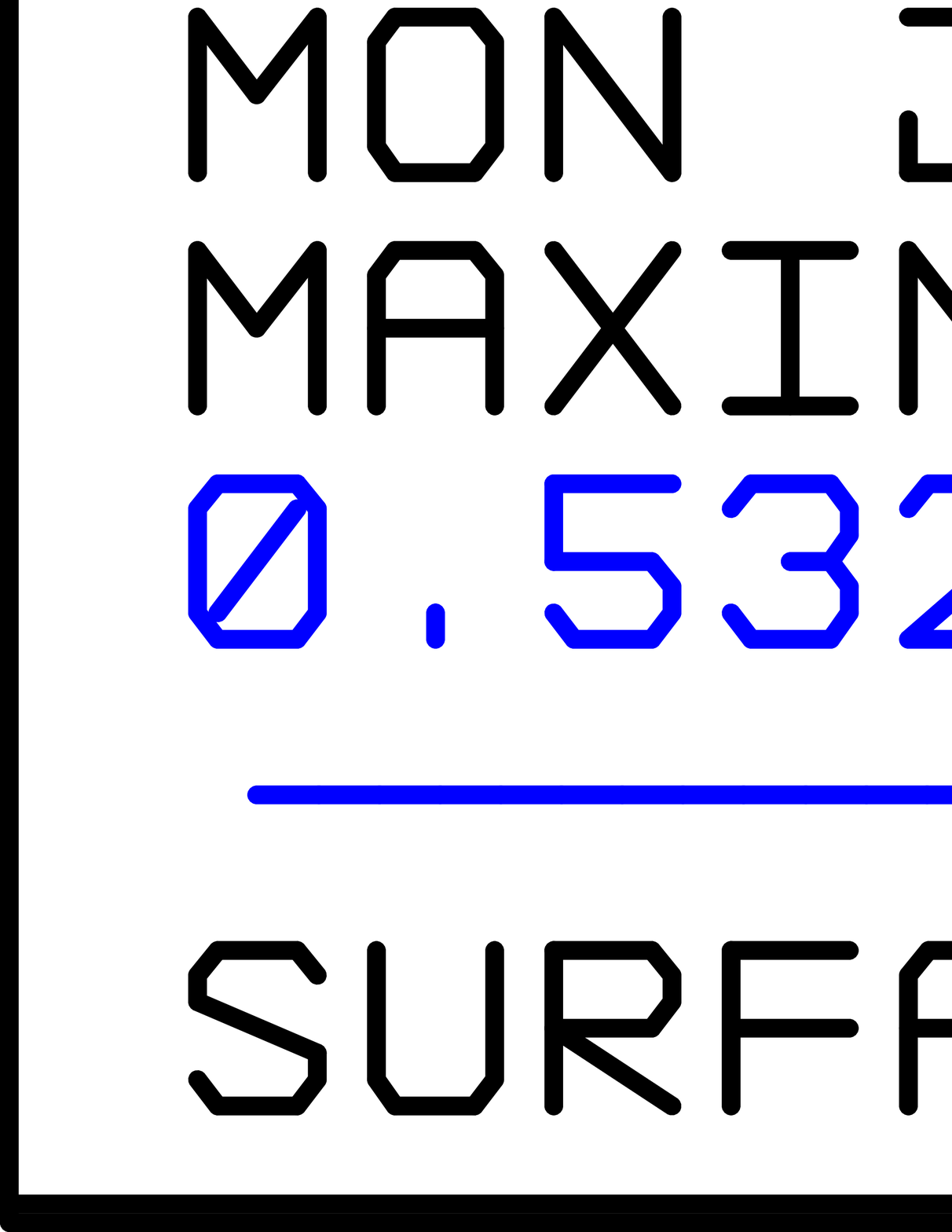}
\end{tabular}
\caption{Aberrations of the F\#/2 lens designed for the ATF extraction line laser-wire: Optical path difference on the left and transverse ray fan on the right. On each figure the 4 pairs of plots correspond to the 4 possible laser beam tilt (0.0, 0.1, 0.2 and 0.3 degrees) for vertical rays (left plot) and horizontal rays (right plot), the three rays in each plot correspond to the 3 different wavelengths studied: 531nm (green), 532nm (blue) and 533nm (red). Each plot give the aberration (vertical axis) as a function of the position of the ray (horizontal axis).
The horizontal axis is normalised to the size of the lens: a ray on the edge of the lens will be on the edge of the plot whereas a central ray will be in the centre of the plot. The scale of the vertical axis is $\pm 5$ waves for the OPD plot and $\pm 5 \mu m$ for the transverse ray fan plot. \label{fig:lens_aberrations} }
\end{center}
\end{figure}

The optimisation process has also been used to reduce the aberrations. The figure~\ref{fig:lens_aberrations} shows the aberrations of the final design.

\begin{figure}[htbp]
\begin{center}
\includegraphics[height=6cm]{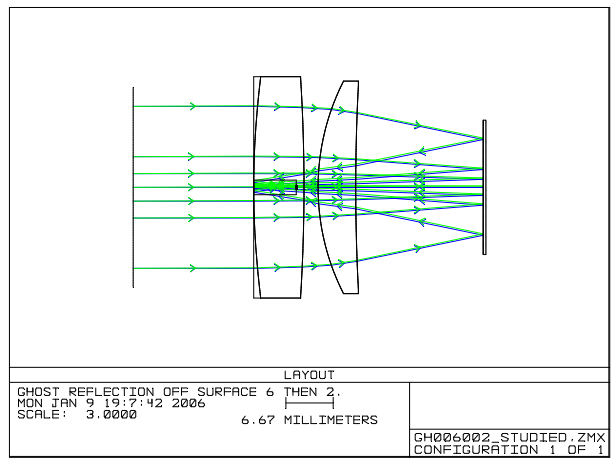}
\caption{Ghost in the lens. This ghost reflection of the lens focus is formed by a reflection on the outer surface of the vacuum window and on the aspherical surface. \label{fig:lens_ghost} }
\end{center}
\end{figure}

Care has been taken to minimise the effect of possible ghosts reflections of the focus in the glass elements. No single bounce ghost remains in the lens but one second order ghost with a radius of $250 \mu m$ remains as shown in figure~\ref{fig:lens_ghost}. As a very low reflectivity coating is being used for the lens,  calculations\cite{lens_calculations} have shown that this ghost should not be an issue.

This F\#/2 lens will later be replaced by a  F\#/1 lens that is currently being designed.

\section{Interaction chamber}
The interaction chamber\cite{bdir_dh} in which the collisions between the laser photons and the electrons will take place has been manufactured in Oxford and is now installed at KEK.

\begin{figure}[htbp]
\begin{center}
\begin{tabular}{ccc}
\includegraphics[height=6cm]{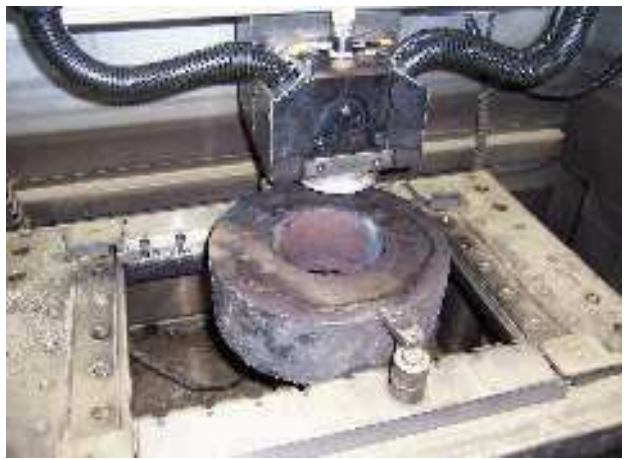} &  & \includegraphics[height=6cm]{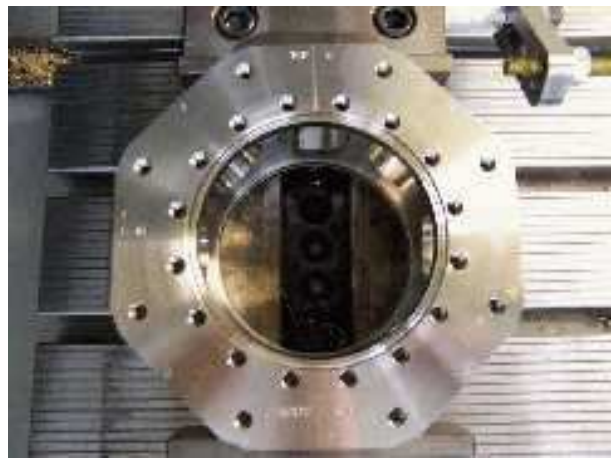}
\end{tabular}
\caption{Raw block of stainless steel used to build the interaction chamber before processing (left) and the same after processing, the finished interaction chamber (right). \label{fig:chamber}  }
\end{center}
\end{figure}

\begin{figure}[htbp]
\begin{center}
\includegraphics[height=18cm,angle=270]{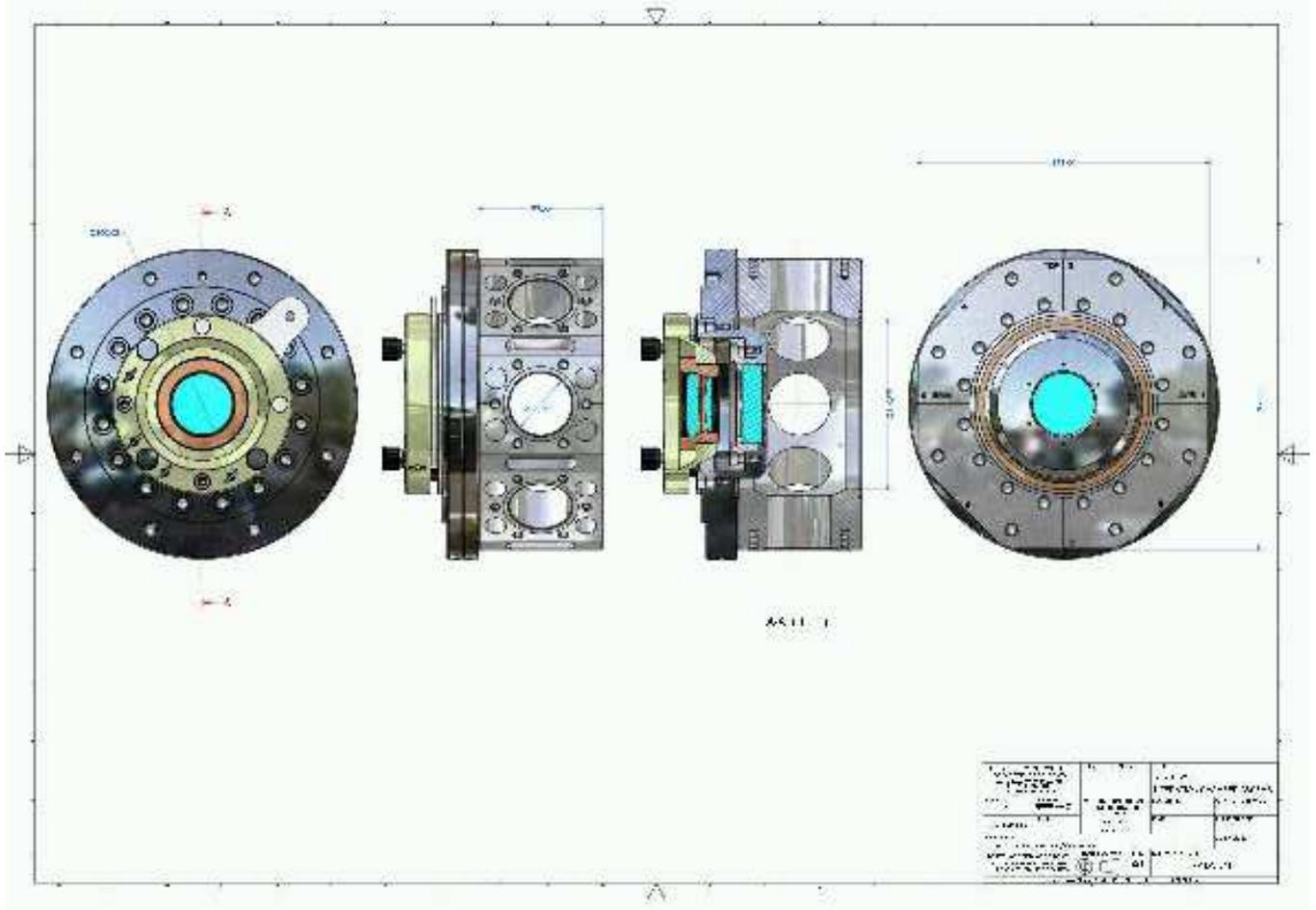}
\caption{Drawings of the ATF extraction line laser-wire interaction chamber. In the leftmost and rightmost drawings the electron beam direction is in the plan of the paper and the laser beam direction is perpendicular to the paper's plan. In the two inner drawings the laser beam is in the plan of the paper and the electron beam is perpendicular to the plan of the paper. \label{fig:chamber_drawing}  }
\end{center}
\end{figure}

Photographs of this chamber are shown in figure~\ref{fig:chamber} and drawings are shown in figure~\ref{fig:chamber_drawing}. The chamber has two ports on which a flange with a window can be mounted. These windows, made of fused silica, are used to allow the laser light to enter the ultra high vacuum of the beam pipe .  
 The tight sealing of the glass on the stainless steel flange is achieved thanks to a novel indium seal designed specially for this purpose. The aperture of these windows is wide enough to allow the use of an F\#/1 lens.
Two other ports of the chamber are used to connect the chamber to the accelerator beam pipe. The chamber has 6 other ports on which diagnostics tools such as a wire-scanner, a screen or a knife edge can be installed.

\section{Conclusion}

The ATF extraction line laser-wire will allow us to demonstrate that laser-wires can be used to measure the very small beams produced in the ILC linac. In the near future we hope that our laser-wire will be an important diagnostic tool at the ATF2.

\section{Acknowledgements}

We would like to thank the ATF group at KEK for their kind support and their availability to our requests.

Work supported in part by the Daiwa Anglo-Japanese Foudation, The Royal Society,
the PPARC LC-ABD Collaboration, and by the Commission of European
Communities under the 6th Framework Programme Structuring the European
Research Area, contract number RIDS-011899.

\bibliographystyle{myunsrt}
\bibliography{biblio}

\end{document}